\documentclass[a4paper,final]{article}
\usepackage{latexsym,amsmath,amssymb,showkeys}

\newcommand{\Mt}{\widetilde{M}} 
\newcommand{\gt}{\tilde{g}}
\newcommand{\scri}{\mathcal{J}}
\newcommand{\RR}{\mathbf{R}}

\newcommand{\cS}{\mathcal{S}}
\newcommand{\cT}{\mathcal{T}}

\newcommand{\eps}{\epsilon}
\newcommand{\del}{\partial}
\newcommand{\2}{\frac12}

\numberwithin{equation}{section}

\title {Numerical treatment of the hyperboloidal initial value problem
for the vacuum Einstein equations\\
I. The conformal field equations}
\author{J\"org Frauendiener\\
Institut f\"ur Theoretische Astrophysik,\\
Universit\"at T\"ubingen,\\
Auf der Morgenstelle 10,\\
D-72076 T\"ubingen,\\
Germany}

\begin{document}
\maketitle

\begin{abstract}
  This is the first in a series of articles on the numerical solution
  of Friedrich's conformal field equations for Einstein's theory of
  gravity. We will discuss in this paper why one should be interested
  in applying the conformal method to physical problems and why
  there is good hope that this might even be a good idea from the
  numerical point of view. We describe in detail the derivation of the
  conformal field equations in the spinor formalism which we use
  for the implementation of the equations, and present all the
  equations as a reference for future work. Finally, we discuss the
  implications of the assumptions of a continuous symmetry.
\end{abstract}

\section{Introduction}

Much of the current work in numerical and experimental relativity is
devoted to obtaining information about the gravitational radiation
that is emitted by astrophysical processes which are taking place in
our universe. The goal is to obtain wave forms, i.e., the ``finger
prints'' by which different processes can be identified when the
gravitational waves are detected by laser interferometers like LIGO or
VIRGO. The common way to describe such a system within Einstein's
theory of gravity is by way of an idealization where the system is
considered as being so far away from the rest of the universe that the
influence of the latter can be neglected (cf. \cite{Friedrich-1996-2}
for a clear discussion of what is involved in this process). Then,
intuitively, the fields far away from the source should decay so that
the space-time becomes asymptotically flat. The detectors are then
idealized as observers which are located ``at infinity'' where they
can gather the gravitational radiation coming from the system.

Isolated gravitating systems and the structure of their ``far fields''
have been investigated for a long time because of their importance for
the interpretation of measurements. A series of articles which heavily
influenced the way we look at the subject today was published in the
early 60's. In these articles various important contributions were
made: the ``peeling property'' of the Weyl tensor \cite{Sachs-1961},
the idea of analysing the vacuum field equations on outgoing null
hypersurfaces resulting in the Bondi mass loss formula
\cite{BondivanderBurgMetzner-1962,Sachs-1962}, the invention of the NP
formalism and the proposal for considering the vacuum Bianchi identity
as a field equation for the Weyl tensor \cite{NewmanPenrose-1962} and
the asymptotic solution of the Einstein vacuum equations
\cite{NewmanUnti-1962}. Assuming that certain components of the Weyl
curvature fall-off in a specific way it was found by formal power
series analysis of the asymptotic characteristic initial value problem
that the fall-off behaviour of the fields along null directions could
be characterized in terms of certain special coordinate systems whose
existence was presupposed.  Finally, it was realized by Penrose
\cite{Penrose-1963} that these fall-off conditions as well as the
peeling property could be understood in a purely geometric way. He
introduced the notion of a conformal extension by which a Lorentz
manifold $(\Mt,\gt)$ is embedded into a bigger manifold $(M,g)$ with
isomorphic conformal structure but with a Lorentz metric which differs
from $\gt$ by a positive factor $g=\Omega^2\gt$. The idea was to study
the global conformal properties of Minkowski space in order to obtain
a criterion for what one should call an ``asymptotically flat''
space-time. Guided by the Minkowski situation Penrose suggested that
such space-times allow the attachment of a conformal boundary $\scri$
which is characterized by the vanishing of the conformal factor
$\Omega$. This boundary is a regular null hypersurface in the ambient
unphysical manifold. It can be interpreted as the points which are at
infinity for the physical manifold along null directions.

The question arises as to what extent this geometric picture is
compatible with the Einstein equations. Friedrich could derive a
system of equations \cite{Friedrich-1981}, the ``conformal field
equations'', which are defined on that larger unphysical
manifold. Furthermore, a solution of the conformal field equations
gives rise to a solution of the standard field equations on the
physical space-time. This system is written in terms of geometric
quantities of the unphysical manifold and the conformal factor
$\Omega$ and it is regular everywhere even at points where $\Omega$
vanishes. In a usual $3+1$ decomposition, the conformal field
equations split into constraint equations and evolution equations.
Using this system of equations Friedrich was able to reduce the
asymptotic characteristic initial value problem for the Einstein
equations where data are given on a part of (past) null infinity and
an ingoing null hypersurface which intersects null infinity in a
two-dimensional surface to a characteristic initial value problem for
a symmetric hyperbolic system \cite{Friedrich-1981}.

In order to describe a physical situation one would like to prescribe
initial data for the conformal field equations on some initial
space-like hypersurface and determine from them the future of the
system. Ideally, the data should be given on an asymptotically flat
space-like surface. It does turn out that the initial data for the
conformal field equations on such a hypersurface are necessarily
singular because the conformal structure of space-time is singular at
$i^0$ whenever the ADM mass is non-zero (see \cite{Friedrich-1996-3}
for a new approach towards the solution of this problem). Therefore,
the initial data are given on a space-like hypersurface which
intersects $\scri$ transversely in a two-dimensional surface. Such
hypersurfaces are called hyperboloidal surfaces because they behave
like spaces of constant negative curvature in the neighbourhood of
$\scri$. Friedrich \cite{Friedrich-1983} has shown that the Cauchy
problem for data given on such hypersurfaces, the hyperboloidal
initial value problem, is well posed, i.e., given smooth initial data
which solve the constraints then there exists a solution of the
evolution equations in some neighbourhood of the initial surface. If
the data are close enough to Minkowski data, then the future
development is complete in the sense that there exists a regular point
$i^+$ whose past light cone coincides with $\scri$.

The fact that the domain of dependence of the initial hyperboloidal
hypersurface includes the complete physical space-time in the future
allows the study of global phenomena like the behaviour of horizons
and the causal structure of singularities. But also, since one has
access to null infinity where gravitational radiation is registered
one can in principle ``extract'' the radiative information by purely
local manipulations from the fields ``on $\scri$''. This suggests that
the hyperboloidal initial value problem is the appropriate device for
examining these issues.  The goal of the present work is the
investigation whether the conformal field equations, and in particular
the hyperboloidal initial value problem, can provide an effective
numerical tool for analysing the global structure of asymptotically
flat space-times and for obtaining information about the gravitational
radiation emitted by the system in question. The starting point for
this investigation was the work by H\"ubner \cite{Huebner-1996} who
was able to demonstrate the feasibility of that approach in the
spherically symmetric case of gravity coupled to a scalar field.

There are various other numerical approaches towards these problems
based on the numerical solution of the standard field equations,
either in terms of a Cauchy problem, a characteristic initial value
problem or a combination of both. The first two alternatives both have
some problems. The standard Cauchy problem cannot provide complete
global information because one has to cut off the initial data surface
and provide boundary data on a time-like hypersurface which intersects
the initial surface in a two-dimensional surface. The boundary data
change the solution in their domain of influence and hence, if the
boundary data are unphysical, so will be the solution. Even if the
boundary conditions are physical, the radiation data obtained are
still only approximate because the boundary is not at infinity. Only
there, radiation can unambigously be defined. Therefore, as a matter
of principle, the standard Cauchy problem can provide only approximate
radiation information. In the hyperboloidal problem there is only one
idealization involved namely that of how an isolated system is to be
described.

The characteristic initial value problem, on the other hand, can be
put to good use in the neighbourhood of $\scri$. Space-time is
foliated by outgoing null hypersurfaces and one can perform a
conformal transformation to obtain a problem where null infinity is at
finite places. This is one boundary for the outgoing initial null
hypersurface. The other boundary is located someplace not too far in
the interior of the space-time.  The numerical procedure for solving
the characteristic initial value problem is relatively simple compared
to the Cauchy problem which is due to the fact that the equations
split into hypersurface equations, which are essentially ordinary
differential equations and one evolution equation.  Three further
equations, the so called ``conservation equations'', have to be
satisfied on $\scri$. The problem with this approach is the fact that
null hypersurfaces invariably tend to form caustics, places where the
hypersurface intersects itself so that it becomes impossible (or at
least very difficult) to give unambiguous initial data. The stronger
the fields are the earlier the caustics will appear.

The last alternative, commonly called the Cauchy-Characteristic
Matching (CCM) procedure (probably going back to \cite{Bishop-1993}),
has been intensely studied by various groups
cf. \cite{BishopEtAl-1997,ClarkedInverno-1994,dInvernoVickers-1997}. The
idea is to combine the two previous approaches without their
respective disadvantages. The procedure is roughly to divide the
physical space-time by a time-like world tube $\cT$ and to evolve the
inner part by solving a Cauchy problem. The exterior of $\cT$ is
evolved by solving the characteristic initial value problem based on
outgoing null hypersurfaces connecting the world tube with null
infinity. An initial hypersurface for the combined problem consists of
a space-like hypersurface $\cS$ with a boundary (which indicates the
intersection of $\cS$ and $\cT$) together with the outgoing null
hypersurface emanating from the boundary. Obviously, at the interface
where the initial hypersurface changes its causal character from
space-like to null there is a non-differentiable kink. The great
challenge is to numerically implement the information exchange across
that kink.

When viewed in the unphysical space-time the initial surface for the
CCM procedure intersects $\scri$ in a two-dimensional ``cut''. Now
consider a space-like hypersurface $\cS$ which also goes through that
same cut. It is clear that this surface is a hyperboloidal
hypersurface. Its domain of dependence is the same as that of the
Cauchy-characteristic hypersurface. The region of $\scri$ which can be
described is the same in both cases. One advantage of evolving with
the conformal field equations is certainly the fact that the causal
character of the foliation does not change so that there is no
interface and no need to change the evolution algorithm. Another
advantage is that one can go smoothly through $\scri$ which allows to
keep $\scri$ in the interior of the grid in order to avoid numerical
influences from the grid boundaries. There are more equations to solve
in the case of the conformal field equations than there are in
standard ADM-like formulations. This might be considered as a
drawback. However, recent formulations of the Einstein equations as
hyperbolic systems \cite{AbrahamsAndersonChoquet-BruhatYork-1997}
introduce many additional variables so that the resulting system is
comparable in size to the system of conformal equations. Furthermore,
the quantities in the latter system which are evolved in addition to
the spatial metric and the extrinsic curvature in the standard case
have a geometric meaning (Ricci- and Weyl tensor components) and any
code which aspires to analyse the space-time structure needs to
compute those quantities anyway.

This present article is meant to be the first in a series of papers on
the numerical treatment of the conformal vacuum field equations. In
this paper, we derive the conformal field equations in a formalism
using space spinors. Although this has been done previously
\cite{Friedrich-1991}, we present the equations here in a form
suitable for our immediate purposes, the main reason being to
establish a common notation and for reference. The space spinor
formalism has the advantage that the equations can easily be
decomposed into evolution and constraint parts and that the evolution
part comes out automatically in symmetric hyperbolic
form. Furthermore, the equations can be written in a more compact form
and the possibility of decomposing spinor fields into their
irreducible parts can be used to remove any redundancy from the set of
unknowns.

In the second article \cite{jf-1997-3} we present the numerical
treatment of the evolution part with the additional assumption of a
symmetry and in the third we want to discuss the solution of the
constraint equations. The conventions used throughout this work are
those of Penrose and Rindler \cite{PenroseRindlerII}.

\section{The conformal field equations}
\label{sec:gendisc}

In this section we want to give a derivation of the conformal field
equations and a brief discussion of their properties. Apart from
introducing the necessary background on the hyperboloidal initial
value problem this section also serves as a reference to the actual
equations used in the code.

An essential ingredient in this approach towards the examination of
global properties of space-times is the notion of a conformal
transformation. Let $(\Mt,\gt)$ be a Lorentz manifold with vanishing
Einstein tensor (we will assume throughout that the cosmological
constant vanishes). Assume that this ``physical space-time'' is such
that the following conditions hold
\begin{itemize}
\item there exists a Lorentz manifold $(M,g)$ with boundary $\scri$
  and a function $\Omega$ on $M$ such that $\Omega \ge 0$ on $M$ and
  $\Omega=0$, $d\Omega\ne0$ on $\scri$,
\item the physical manifold can be identified with the interior of $M$
  and there the equation $g=\Omega^2\gt$ holds.
\end{itemize}
These conditions state that the physical manifold is conformal to the
interior of the ``unphysical'' manifold $M$. The points on the
boundary $\scri$ can be thought of as representing the points of $\Mt$
which are ``at infinity'' with respect to the physical metric
$\gt$. With the vanishing of the cosmological constant it follows that
$\scri$ is a regular null hypersurface in $M$ on which the Weyl
curvature vanishes (although this is only strictly proven in the case
where $\scri$ has the topology $S^2\times\RR$ \cite{Penrose-1965}).

The conformal field equations can now be obtained from the basic
geometric equations on $M$ and $\Mt$, the Einstein equation which
holds on $\Mt$ and the conformal transformation properties of the
geometric fields. In view of the numerical application it is
advantageous to have a first order system. This is easily achieved by
using a frame formalism. The system then consists of the following
equations: 
\begin{itemize}
\item The first of Cartan's structure equations, which expresses the
  fact that the connection on $M$ is torsion free. It can be viewed as
  an equation for the components of the chosen tetrad with respect to
  the chosen coordinates.
\item The second structure equation which relates the Ricci rotation
  coefficients of the connection to the curvature components. It can
  be viewed as an equation for the connection components with respect
  to the chosen tetrad.
\item The  Bianchi identity for $\gt$. This is an identity which
  relates the derivatives of the physical Ricci and the Weyl
  curvature. Since $\Mt$ is a vacuum space-time this yields an
  equation for the physical Weyl curvature. Expressing this equation
  in terms of the unphysical connection yields an equation for the 
  rescaled Weyl curvature $D_{abcd}=\Omega^{-1} C_{abcd}$, which looks
  formally like the familiar spin-2 zero rest-mass equation. 
\item The Bianchi identity for $g$. Again, this is a relation between
  the derivatives of the Ricci and the Weyl curvature, but now on the
  unphysical space-time. Using the equation of the rescaled Weyl
  curvature this identy yields an equation for the unphysical Ricci
  curvature. 
\item equations for the conformal factor $\Omega$ and its derivatives
  obtained from the conformal transformation law for the Ricci
  curvature, and
\item an equation for the function $S:=\frac14\Box\Omega$, which is a
  consequence of the earlier equations.
\end{itemize}
Due to the geometric origin of these equation, there is gauge freedom in
this system. Several variables can be chosen freely. Apart from the
coordinates, this is true also for the tetrad which is fixed by the
metric only up to Lorentz transformations and for the conformal
factor, which is fixed by the conditions above up to multiplication
with a strictly positive function $\Omega\mapsto \theta \Omega$, where
$\theta >0$ on $M$. This allows the free choice of eleven functions,
the gauge source functions, which can be fixed in numerous ways. It is
here where the development of a code to evolve space-times turns into
an art.

The essential property of the above system is the following: With the
gauge source functions fixed as arbitrary functions of the
coordinates, the system can be decomposed by a usual $3+1$ splitting
into two separate systems with respect to a givem foliation of
space-like hypersurfaces. The first of those, the constraints, is
intrinsic to the space-like hypersurfaces and therefore it restricts
the values of the variables there. The second part, the evolution
equations, can be written as a quasi-linear symmetric hyperbolic
system. This has the consequence that the Cauchy problem for this
system is well posed: given initial data for the unknown functions on
a space-like hypersurface $\cS$ then in a neighbourhood of $\cS$ there
will exist a solution of the system acquiring the prescribed values on
$\cS$. It turns out, that once the constraints are satisfied on the
initial hypersurface they will be satisfied everywhere by virtue of
the evolution equations, they are propagated by the
evolution. Therefore, given initial data which satisfy the constraints
then they will evolve into a solution of the conformal field
equations.

If $M$ is such that the initial hypersurface $\cS$ and $\scri$
intersect transversely in a regular compact two-dimensional surface
then one can talk about the hyperboloidal initial value problem. The
standard example for such ``hyperboloidal'' surfaces are the conformal
images of the space-like hyperboloids in Minkowski space in the usual
conformal picture. The fact that $\cS$ and $\scri$ intersect in the
unphysical space-time means in physical terms that the space-like
surface extends out to null infinity. Thus, such surfaces are not
Cauchy surfaces for the standard Cauchy problem for the Einstein
equations in the physical space-time.

In summary, the conformal field equations allow a well-posed initial
value problem on space-like hypersurfaces in the unphysical space-time
whose physical ``pre-images'' extend asymptotically towards null
infinity. Initial data which satisfy the constraints evolve into a
solution of the complete system.

\section{Space spinors}
\label{sec:spsp}

In this section we briefly introduce the basic formalism used to write
down the equations and to separate them into the evolution and
constraint part. This can very conveniently be achieved using the
space spinor formalism \cite{Sommers-1980}, which in addition allows
writing (and coding) the equations in a more compact form.

The essential ingredient in the space spinor formalism is a time-like
vector field $t^a$ which is normalised by $t_at^a=2$ (note, that we
use throughout the conventions of \cite{PenroseRindlerII}). In terms of
spinors, we have $t^a=t^{AA'}$ and $t_{AA'} t^{BA'} = \eps_A{}^B$. The
existence of this vector field allows the conversion of all primed
spinor indices to unprimed ones by extension of the map $\pi_{A'}
\mapsto \pi_{A'} t^{A'}{}_A$ to the full spinor algebra. E.g., any
covector $v_a=v_{AA'}$ is mapped to $v_{AB}=v_{AA'}t^{A'}{}_B$. Note,
that this spinor can be decomposed into irreducible parts: $v_{AB}=1/2\,
\eps_{AB} v + \tilde{v}_{AB}$ where $\tilde{v}_{AB} =
\tilde{v}_{BA}$. In terms of the original covector these parts
correspond to the components along $t^a$, $v=t^av_a$, and orthogonal to
to $t^a$. 

The vector $t^{AA'}$ can be used to define a complex conjugation map
on the algebra of unprimed spinors by extension of the map $\pi_A
\mapsto \hat{\pi}_A:= t_A{}^{A'}\bar\pi_{A'}$ to the full
algebra. Note, that $\widehat{\hat{\pi}}_{A_1\ldots A_n} = (-1)^{n}
\pi_{A_1\ldots A_n}$. An even valence spinor $\pi_{A_1\ldots A_{2n}}$
is called real, iff $\hat{\pi}_{A_1\ldots A_{2n}} = (-1)^{n}
\pi_{A_1\ldots A_{2n}}$. 

The derivative operator $\nabla$ on $M$ can be decomposed as follows
\begin{eqnarray}
  \label{eq:nabla}
  \nabla_{AA'} &=& \frac{1}{2}\; t_{AA'} D - t_{A'}{}^B D_{AB},\\
\noalign{or, equivalently}
   t^{A'}{}_B \nabla_{AA'} &=& \frac{1}{2}\; \eps_{AB} D + D_{AB},
\end{eqnarray}
where $D:=t^a\nabla_a$ and $D_{AB}=t^{A'}{}_{(B} \nabla_{A)A'}$ are the
parts which act along and perpendicular to $t^a$, respectively.  Thus,
the general procedure we will follow is to write the equations in
spinorial form, then convert to space spinors and finally decompose
them into irreducible parts. 

The derivative of $t^{AA'}$ gives rise to two important spinor fields,
$K_{AB} = t^{A'}{}_B Dt_{AA'}$ and $K_{ABCD} = t^{C'}{}_D D_{AB}
t_{CC'}$. Note the symmetry and reality properties of these fields:
$K_{AB} = K_{(AB)} = -\hat{K}_{AB}$ and $K_{ABCD} = K_{(AB)(CD)} =
\hat{K}_{ABCD}$. Geometrically, $K_{AB}$ corresponds to the
acceleration vector of $t^a$ while $K_{ABCD}$ is related to the
geometry of the distribution defined by $t_a=0$. This distribution is
integrable if and only if $K^A{}_{(BD)A}=0$. Then $t^a$ is hypersurface
orthogonal and $K_{ABCD}$ corresponds to the extrinsic curvature of
the orthogonal surfaces.

We will assume henceforth that $t^a$ is hypersurface
orthogonal. Hence, the covector $t_a$ is proportional to the conormal
of the space-like hypersurfaces given by $t_a=0$. Then, the derivative
$D_{AB}$ is the so called Sen-Witten connection which plays an
important r\^ole in various areas of general relativity. It is not
completely intrinsic to the hypersurfaces but contains information
about the embedding of the surfaces in space-time. This is reflected
in the fact that the connection thus defined possesses torsion which
is proportional to the extrinsic curvature of the surfaces. Therefore,
to obtain a completely intrinsic covariant derivative operator on the
hypersurfaces we define for an arbitrary spinor $\pi_C$ the operator
\begin{equation}
  \label{eq:dab}
  \del_{AB} \pi_C:= D_{AB} \pi_C + \2 K_{ABC}{}^D \pi_D.
\end{equation}
The connection defined by this derivative operator is torsion free and
respects the intrinsic metric of the hypersurfaces. Thus, it is the
$SU(2)$ spin connection of the intrinsic metric. In complete analogy,
we define the operator
\begin{equation}
  \label{eq:d}
  \del \pi_C:= D \pi_C + \2 K_{C}{}^D \pi_D.
\end{equation}
This operator defines a connection along the integral curves of $t^a$
which turns out to be the spinorial equivalent of the Fermi-Walker
connection. E.g., a vector $v^a=v^{AB}$ is Fermi-Walker transported
along the $t^a$ curves iff $\del v^{AB}=0$. 

These two operators are real in the sense that they map real fields
into real fields, which is obvious from the relations
\begin{eqnarray}
  \label{eq:dconj}
  \widehat{\del \pi_C}      &=&  \del \hat\pi_C, \\
  \widehat{\del_{AB} \pi_C} &=& -\del_{AB} \hat\pi_C.
\end{eqnarray}

In order to phrase the structure equations in this formulation we need
to know about the commutators of these operators because these define
the torsion and the curvature. The commutators are given by the
formula, valid for an arbitrary spinor $\alpha_C$
\begin{multline}
  \label{eq:comm1}
  \left[ \del, \del_{AB}\right]\alpha_C = 
  \2 K_{AB} \del \alpha_C + K_{AB}{}^{EF} \del_{EF}\alpha_C
  -\Box_{AB}\alpha_C + \widehat{\Box}_{AB}\alpha_C \\
  +{}\2\left\{\del_{AB} K_{CD} - \del K_{ABCD} - K_{AB}{}^{EF}
    K_{CDEF} \phantom{\2} \right. \\
\left.
    + K_{(C}{}^E K_{D)EAB} + \2 K_{AB} K_{CD}  \right\} \alpha^D
\end{multline}
and
\begin{multline}
  \label{eq:comm2}
  2\del_{E(A} \del^E{}_{B)} \alpha_C = \Box_{AB}\alpha_C +
  \widehat{\Box}_{AB}\alpha_C \\
  - \left\{
         \del_{E(A} K^E{}_{B)CD} 
         - \2 K K_{ABCD} 
         + \2 K_{EFAB} K^{EF}{}_{CD} \right. \\
         \left. + \frac14 \eps_{C(A}\eps_{B)D} 
                  \left(K_{EFGH} K^{EFGH} -  K^2\right)
    \right\} \alpha^D,
\end{multline}
where we have defined the trace of the extrinsic curvature $K :=
K^{AB}{}_{AB}$ and introduced the curvature derivations
\begin{align}
  \label{eq:curvderiv}
  \Box_{AB} \alpha_C &= - \Psi_{ABC}{}^D \alpha_D + 2 \Lambda
  \eps_{C(A} \alpha_{B)}, \\
  \widehat{\Box}_{AB} \alpha_C &= - \widehat\Phi_{ABC}{}^D \alpha_D.
\end{align}
Here, we have used the curvature spinors, i.e., the Weyl spinor
$\Psi_{ABCD}$, the (space spinor equivalent of the) Ricci spinor
$\Phi_{ABCD}:=\Phi_{ABA'B'}t^{A'}{}_Ct^{B'}{}_D$ and the scalar
curvature $\Lambda$.

In view of the reality properties of the derivative operators we find
that the commutation relation can be split into various parts. This
yields equations for the spinor fields $K_{AB}$ and $K_{ABCD}$ 
\begin{gather}
  \label{eq:Keqns}
  \begin{split}  
    \del &K_{ABCD} = \2 \left(\del_{AB} K_{CD} + \del_{CD} K_{AB}
    \right) +K_{AB}{}^{EF} K_{EFCD} + \2 K_{AB} K_{CD} \\ 
  &+ \left(\Phi_{ABCD} + \Phi_{CDAB} \right)
  - \left(\widehat\Psi_{ABCD} + \Psi_{ABCD}\right)
  + 4 \Lambda \eps_{C(A}\eps_{B)D},
  \end{split} \\
  \del_{C(A} K^C{}_{B)} = 0, \\
  \del_{E(A} K^E{}_{B)CD} = 
  \2 \left(\Psi_{ABCD} - \widehat\Psi_{ABCD}\right)
  - \2 \left(\Phi_{ABCD} - \Phi_{CDAB} \right)
\end{gather}
and simplified commutation relations
\begin{multline}
  \left[ \del, \del_{AB} \right] \alpha_C = 
  \2 K_{AB} \del\alpha_C + K_{AB}{}^{EF} \del_{EF} \alpha_C
  + \2 K^E{}_{(C} K_{D)EAB} \alpha^D \\
  - \2 \left(\Psi_{ABCD} - \widehat\Psi_{ABCD}\right) \alpha^D
  - \2 \left(\Phi_{ABCD} - \Phi_{CDAB} \right) \alpha^D
\end{multline}
\begin{multline}
  2 \del_{E(A} \del^E{}_{B)} \alpha_C = \\
  \left\{
    \2 K K_{ABCD} - \2 K_{EFAB} K^{EF}{}_{CD} 
    - \frac14 \eps_{C(A} \eps_{B)D} \left( K^{EFGH} K_{EFGH}
      -K^2\right)
  \right\} \alpha^D \\
  - 2 \Lambda  \eps_{C(A} \eps_{B)D} \alpha^D
  + \2 \left(\Psi_{ABCD} + \widehat\Psi_{ABCD}\right) \alpha^D
  + \2 \left(\Phi_{ABCD} + \Phi_{CDAB} \right) \alpha^D
\end{multline}
When acting on functions, these commutators yield
\begin{gather}
  \left[ \del, \del_{AB} \right] f = 
  \2 K_{AB} \del f + K_{AB}{}^{EF} \del_{EF} f,\\
  2 \del_{E(A} \del^E{}_{B)} f = 0.
\end{gather}

\section{The equations}
\label{sec:eqns}

We are now in a position to give the derivation of the conformal field
equations on the unphysical manifold $M$. To this end we need to
introduce coordinates and a spin frame with respect to which we
express all the spinor fields involved. Since we have already assumed
the existence of a family of space-like hypersurfaces we may now
introduce a time coordinate $t$ be requiring that it be constant with
non vanishing differential on the hypersurfaces. Then, necessarily, we
have $t_a\propto \nabla_a t$. Now, we choose arbitrary coordinates
$\{x^1,x^2,x^3\}$ on $M$. These coordinates can be characterised by
their change along the integral curves of $t^a$, thus defining lapse
function and shift vector.

The choice of frame is probably best described in terms of orthonormal
tetrads. On the initial surface we choose the time-like leg of the
tetrad to be proportional to $t^a$. Then the other members are
tangent to the surface and we choose them arbitrarily. To propagate
the tetrad off the hypersurface, we could use the Fermi-Walker
transport along the integral curves of $t^a$ which has the property
that it leaves the angles between vectors constant, while keeping the
tangent vector along the curve fixed. However, this is not the most
general propagation law with these properties. Any other one with the
above properties differs from the Fermi-Walker transport law by the
addition of a term which involves an infinitesimal rotation. Thus, we
fix an arbitrary transport law which leaves angles invariant and fixes
the tangent vector to propagate the frame into the full space-time. The
infinitesimal rotation involved in the transport law determines some
of the Ricci rotation coefficients of the tetrad thus obtained.

In terms of spinors this choice of frame is expressed as follows. On
the initial surface we choose a normalised spinor field $o_A$, i.e.,
we have $\hat{o}^Ao_A=1$. To complete the spin frame we define
$\iota_A = \hat{o}_A$. Then the orthonormal tetrad constructed from
this spin frame has the above properties on the initial surface. Now
we impose the transport equation in the form 
\begin{equation}
  \label{eq:fw}
  \del o_A = F_{AB} o^B,
\end{equation}
where $F_{AB}$ is an arbitrary purely imaginary and symmetric spinor
field. It corresponds exactly to the infinitesimal rotation mentioned
above. It was already pointed out that $\del$ corresponds to the
Fermi-Walker transport, which is, consequently, selected by choosing
$F_{AB}=0$.

The connection is defined by the spinor fields $K_{AB}$ and $K_{ABCD}$,
the field $F_{AB}$ and finally by the field $\Gamma_{ABCD}$, defined
by
\begin{equation}
  \label{eq:Gamma}
  \del_{AB} o_C = \Gamma_{ABCD} o^D.
\end{equation}
It satisfies the relations $\Gamma_{ABCD} = \Gamma_{(AB)(CD)} =
-\hat\Gamma_{ABCD}$, thus being purely imaginary.

The frame components with respect to a coordinate basis are usually
obtained by applying the vector fields which make up the tetrad to the
coordinates. Similarly, in the present case: we apply the operators
$\del$ and $\del_{AB}$ to the coordinates
\begin{align}
  \label{eq:frame}
  \del t &= \frac1N, & \del_{AB} t &= 0, \\
  \del x^i &= -T^i, & \del_{AB} x^i &= C^i_{AB}.
\end{align}
This defines several additional fields on $M$ which fix the frame in
terms of the chosen coordinates. The second equation reflects the fact
that we have chosen the time-like leg of the tetrad to be the unit
normal of the surfaces throughout. The function $N$ is the lapse while
the three functions $T^i$ are closely related to the shift vector
which appear in all variants of $3+1$ decompositions. 

To make the relationship between the frame as defined above and the
coordinate basis somewhat more precise let us introduce the 1-forms
$\theta$, $\theta^{AB}$ which are dual to the operators $\del$,
$\del_{AB}$ considered as vector fields on $M$. I.e. for any spinor
field $\alpha^{AB}$ we have the relations
\begin{align}
  \langle \theta, \alpha^{AB}\del_{AB} \rangle &= 0, &
  \langle \theta^{AB}, \alpha^{CD}\del_{CD} \rangle &= \alpha^{AB}, \\
  \langle \theta, \del \rangle &= 1, &
  \langle \theta^{AB}, \del \rangle &= 0.
\end{align}
Then the metric $g$ on $M$ when expressed in terms of the present
formalism is simply
\begin{equation}
  g=2\theta\otimes\theta - \theta_{AB} \otimes \theta^{AB}.
\end{equation}
We have the following relations between the coordinate differentials
and the forms $\theta$, $\theta^{AB}$:
\begin{align}
  dt &= \frac1N \theta,&
  dx^i &= C^i_{AB} \theta^{AB} - T^i \theta, \\
  \theta &= N dt,&
  \theta^{AB} &= D^{AB}_k\left( dx^k + T^k N dt \right),
\end{align}
with $D^{AB}_k$ being the inverse of $C^i_{AB}$, i.e., $D^{AB}_k
C^i_{AB}=\delta^i_k$. From these we get the metric expressed in terms
of the frame components
\begin{equation}
\label{eq:metric}
  g=2N^2 dt^2 - D^{AB}_i D_{ABk} 
  \left(dx^i + N T^i dt\right) \otimes \left(dx^k + N T^k dt\right)
\end{equation}

We now come to the system of equations. To express the first structure
equation in terms of the fields defined above, we apply the
commutators to the coordinates which gives the equations
\begin{gather}
\label{eq:first}
\del_{AB} N = \2 N K_{AB},  \\
\del_{(A}{}^C K_{B)C} = 0, \\
\del C_{AB}^i + \del_{AB} T^i = -\2 K_{AB} T^i + K_{AB}{}^{EF}
C_{EF}^i, \\
\del^C{}_{(A} C^i_{B)C} = 0.
\end{gather}

The second structure equation is obtained analogously by applying the
commutators to $o_C$ yielding the evolution equation for $\Gamma$
\begin{multline}
  \label{eq:dGamma}
  \del \Gamma_{ABCD} = \del_{AB} F_{CD} 
  -  2 \Gamma_{ABE(D} F^E{}_{C)}
  + \2 K_{AB} F_{CD} \\
  + K_{AB}{}^{EF} \Gamma_{EFCD} 
  + \2 K^E{}_{(C} K_{D)EAB} \\
  - \2 \left\{ \Psi_{ABCD} - \hat\Psi_{ABCD} \right\}
  - \2 \left\{ \Phi_{ABCD} - \Phi_{CDAB} \right\}
\end{multline}
and the constraint equation
\begin{multline}
  \label{eq:gradGamma}
  2 \del_{E(A} \Gamma^E{}_{B)CD} = 
  2 \Gamma^E{}_{(B}{}^F{}_C \Gamma_{A)EDF} \\
  + \2 \left\{
     K K_{ABCD} -  K_{EFAB} K^{EF}{}_{CD} 
    - \2 \eps_{C(A} \eps_{B)D} \left( K^{EFGH} K_{EFGH} - K^2\right)
  \right\} \\
  - 2 \Lambda \eps_{C(A}\eps_{B)D} 
  + \2 \left\{ \Psi_{ABCD} + \hat\Psi_{ABCD} \right\}
  + \2 \left\{ \Phi_{ABCD} + \Phi_{CDAB} \right\}.
\end{multline}

The derivation of the equation for the curvature and the conformal
factor will be given first using primed and unprimed spinors. At the
end of this section we will collect all the equations in the space
spinor formalism. To obtain equations for the curvature components we
need to look at the Bianchi identities. These have the following
spinorial form
\begin{gather}
\label{eq:bianchi}
  \nabla_{B'}{}^A\Psi_{ABCD} = \nabla^{A'}{}_{(B} \Phi_{CD)A'B'} \\
  \nabla^{BB'} \Phi_{ABA'B'} + 3 \nabla_{AA'} \Lambda = 0,
\end{gather}
which is valid in any space-time. Since the Weyl curvature is
conformally invariant the first of these equations when viewed in the
physical space-time takes the form
\begin{equation}
  \tilde{\nabla}_{B'}{}^A\Psi_{ABCD} = 0,
\end{equation}
with $\tilde\nabla$ being the physical connection. Here we have used
the Einstein equation and the fact that the energy-momentum tensor in
the physical space-time vanishes. The conformal transformation
behaviour of the connection implies that this equation is conformally
invariant provided we rescale $\Psi_{ABCD}$ with the conformal factor
in the appropriate way. Thus, if we define the rescaled Weyl spinor
\begin{equation}
  \psi_{ABCD} = \Omega^{-1} \Psi_{ABCD}
\end{equation}
then we have the equation 
\begin{equation}
  \label{eq:psi1}
  \nabla_{B'}{}^A\psi_{ABCD} = 0
\end{equation}
holding on $M$. Note, that $\psi_{ABCD}$ is well behaved on $M$,
because the Weyl curvature vanishes on $\scri$, where also
$\Omega=0$.
 
Expressing the Bianchi identities \eqref{eq:bianchi} in
terms of the rescaled Weyl spinor and using the equation
\eqref{eq:psi1} together with the definition
\begin{equation}
  \label{eq:dOmega}
  \Sigma_{AA'}:=\nabla_{AA'} \Omega
\end{equation}
we get an equation for the Ricci spinor
\begin{equation}
  \label{eq:ricci1}
  \nabla_B{}^{B'} \Phi_{CDA'B'} = 
  \Sigma_{A'}{}^E \psi_{EBCD} + 2 \eps_{B(C} \nabla_{D)A'} \Lambda.
\end{equation}
When these equations are expressed in terms of space spinors using the
operators $\del$ and $\del_{AB}$ they become rather long. To simplify
them we decompose every spinor and every equation into their
irreducible parts thus obtaining a set of equations which can be
further decomposed into real and imaginary parts. 
The resulting equations will be displayed below.

The next set of equations is the definition of $\Sigma$ viewed as an
equation for $\Omega$ and the equation which expresses the conformal
transformation behaviour of the Ricci spinor viewed as an equation for
$\Sigma_{AA'}$
\begin{equation}
  \label{eq:sigma}
  \nabla_{BB'} \Sigma_{AA'} = - \Omega \Phi_{ABA'B'} + \eps_{AB}
  \eps_{A'B'} S
\end{equation}
with $S:=1/4 \Box\Omega$. The final equation is an equation for $S$
which can be derived from the equations established so far. It reads
\begin{equation}
  \label{eq:s}
  \nabla_{AA'} S = - \Phi_{ABA'B'} \Sigma^{BB'} + \Omega \nabla_{AA'}
  \Lambda + 2 \Lambda \Sigma_{AA'}.
\end{equation}
The conformal transformation behaviour of the scalar curvature implies
an algebraic equation
\begin{equation}
  \label{eq:alg}
  2\Omega S - 2 \Omega^2 \Lambda - \Sigma_{AA'} \Sigma^{AA'} = 0.
\end{equation}

This completes the derivation of the system of equations and we now
want to present the full list of equations in the space spinor
formalism. The variables for which we have an evolution equation are:
$N$, $C^i_{AB}$, $K_{AB}$, $K_{ABCD}$, $\Gamma_{ABCD}$,
$E_{ABCD}:=\2\left(\psi_{ABCD} + \hat\psi_{ABCD}\right)$,
$B_{ABCD}:=\frac1{2i}\left(\psi_{ABCD} - \hat\psi_{ABCD}\right)$,
$\phi_{ABCD}$, $\phi_{AB}$, $\phi$, $\Omega$, $\Sigma$, $\Sigma_{AB}$
and $S$. In this list we have included the irreducible parts of
$\Phi_{ABCD}$ and $\Sigma_{AA'}$ defined by the decompositions
\begin{align}
  \label{eq:decomp}
  \Sigma_{AA'} &= \2 t_{AA'} \Sigma - t_{A'}{}^{B} \Sigma_{AB}, \\
  \Phi_{ABCD} &= \phi_{ABCD} + \2 \eps_{A(C} \phi_{D)B} + \2
  \eps_{B(C} \phi_{D)A} - \frac13 \eps_{A(C} \eps_{D)B} \phi.
\end{align}

We have included in this list the lapse $N$ and $K_{AB}$ for which we
do not have an evolution equation yet. This can easily be achieved by
computation of the ``harmonicity function'' $F:=2 \Box t$. Expressing
the wave operator in terms of $\del$ and $\del_{AB}$ yields the
equation for $N$, while commuting $\del$ and $\del_{AB}$ on $N$ gives
the evolution equation for $K_{AB}$.

The evolution equations can be grouped together according to the
geometric meaning of the variables:
\begin{itemize}
\item The evolution of the frame components
\begin{align}
  \del N &= - K N - N^2 F, \label{eq:evN}\\ 
  \del C^i_{AB} &= K_{AB}{}^{EF} C_{EF}^i - \del_{AB} T^i - \2 K_{AB}
  T^i. \label{eq:evC}
\end{align}

\item The evolution of the extrinsic curvature and the acceleration
vector
\begin{multline}
\label{eq:evK2}
  \del K_{AB} + 2 \del^{CD} K_{ABCD} = \\
  K_{AB}{}^{EF} K_{EF} - K_{AB} K - 4 \phi_{AB} -2N\del_{AB}F -N
  K_{AB} F,
\end{multline}
\begin{multline}
\label{eq:evK4}
    \del K_{ABCD} - \2 \left(\del_{AB} K_{CD} + \del_{CD} K_{AB}
    \right) = \\
    K_{AB}{}^{EF} K_{EFCD} + \2 K_{AB} K_{CD} \\ 
  + 2\phi_{ABCD}
  - 2 \Omega E_{ABCD}
  + \frac23\left(6 \Lambda + \phi\right) \eps_{C(A}\eps_{B)D},
\end{multline}

\item The evolution of the intrinsic connection
\begin{multline}
\label{eq:evGamma4}
   \del \Gamma_{ABCD} = \del_{AB} F_{CD} 
  -  2 \Gamma_{ABE(D} F^E{}_{C)}
  + \2 K_{AB} F_{CD} \\
  + K_{AB}{}^{EF} \Gamma_{EFCD} 
  + \2 K^E{}_{(C} K_{D)EAB} \\
  - i \Omega B_{ABCD}
  - \eps_{(A(C}\phi_{D)B)} 
\end{multline}
\item The evolution of the Ricci curvature
  \begin{multline}
    \label{eq:evphi4}
    \del \phi_{ABCD} - \del_{(AB} \phi_{CD)} = \\
    K_{(AB} \phi_{CD)} + K_{(AB}{}^{EF} \phi_{CD)EF} 
    - \frac23 K_{(ABCD)} \phi \\
    - \Sigma E_{ABCD} + 2 i \Sigma_{(A}{}^E B_{BCD)E},
  \end{multline}
  \begin{multline}
\label{eq:evphi2}
    \del \phi_{AB} - \frac23 \del_{AB} \phi + \del^{CD}\phi_{ABCD}= \\
    \frac23 K_{AB} \phi - \phi_{ABCD} K^{CD} + \frac32 K_{ABCD}
    \phi^{CD} - \2 K \phi_{AB} \\
    + E_{ABCD} \Sigma^{CD} - 4 \del_{AB}
    \Lambda,     
  \end{multline}
  \begin{multline}
\label{eq:evphi}
    \del \phi + 2 \del_{AB} \phi^{AB} = \\
    - 2 K_{AB} \phi^{AB} - 2 K^{ABCD} \phi_{ABCD} + \frac43 K\phi - 2
    \del \Lambda
  \end{multline}
\item The evolution of the Weyl curvature
  \begin{multline}
\label{eq:evE4}
    \del E_{ABCD} - 2i \del_{(A}{}^E B_{BCD)E} = \\
     2i K_{(A}{}^E B_{BCD)E} - 3 K_{(AB}{}^{EF} E_{CD)EF} + 2 K E_{ABCD}    
  \end{multline}
  \begin{multline}
\label{eq:evB4}
    \del B_{ABCD} + 2i \del_{(A}{}^E E_{BCD)E} = \\
    -2i K_{(A}{}^E E_{BCD)E} - 3 K_{(AB}{}^{EF} B_{CD)EF} + 2 K B_{ABCD}    
  \end{multline}
\item The evolution of the conformal factor
  \begin{align}
    \del \Omega &= \Sigma, \label{eq:evOmega} \\
    \del \Sigma &= -K_{AB} \Sigma^{AB} - \Omega \phi + 2S,\label{eq:evSigma} \\
    \del \Sigma_{AB} &= \2 K_{AB} \Sigma - \Omega \phi_{AB},
    \label{eq:evSigma2}\\ 
    \del S &= \2 \phi \Sigma - \phi_{AB} \Sigma^{AB} + \Omega
    \del\Lambda + 2 \Lambda \Sigma.\label{eq:evS}
  \end{align}
\end{itemize}
This completes the list of evolution equations. We need to make
several remarks.
\begin{itemize}
\item The gauge functions in this system are the ``harmonicity'' $F$,
  the shift functions $T^i$, the frame rotations $F_{AB}$ and the scalar
  curvature $\Lambda$. These eight functions can be chosen almost at
  will. From the form of the metric we infer a condition which has to be
  satisfied by the shift functions. Since the vector $\del_t$ needs to
  be time-like, we find the inequality 
  \begin{equation}
    \label{eq:T2ge2}
    D^{AB}_k D_{ABi} T^i T^k > 2.    
  \end{equation}
  The full gauge freedom of eleven functions has been reduced to
  eight because of our fixing of the time-like leg of the tetrad.
\item Since the operator $\del$ is the directional derivative
  along the vector field $t^a$ when acting on functions, most of the
  equations are simple advection equations along that vector field. This
  is obvious from the explicit form of $\del f = \frac1N \del_t - T^i
  \del_{x^i}$. There are three subsystems for which this is not the
  case. These are the systems describing the evolution of $K_{AB}$ and
  $K_{ABCD}$, of the Ricci curvature and of the Weyl curvature. They
  will have considerable significance later when it comes to the
  numerical treatment of the equations at the boundary.
\item It is a useful property of the space spinor formalism that the
  equations automatically come out separated into constraints and
  evolution equations and that the evolution equations automatically
  come out as a symmetric hyperbolic system. This is
  the case for the above system. As written it is in symmetric
  hyperbolic form. The symmetric hyperbolicity is the basic property
  which allows the proof of existence and uniqueness of solutions for
  various initial value problems see e.g.\cite{Friedrich-1983}.
\end{itemize}

In a similar way, the constraint equations can be grouped according to
their geometric meaning starting with
\begin{itemize}
\item the frame components
  \begin{align}
    0 &= \del_{AB} N - \2 K_{AB} N,\label{eq:cnN}\\
    0 &= \del_{(A}{}^{C} C^i_{B)C}.\label{eq:cnC}
  \end{align}
\item Next are the extrinsic curvature and the acceleration of the
  time-like unit normal to the surfaces
  \begin{align}
    0 &= \del_{(A}{}^{C} K_{B)C},\label{eq:cnK2}\\
    0 &= \del_{(A}{}^{E} K_{B)ECD} + i \Omega B_{ABCD} 
       - \eps_{(A(C}\phi_{D)B)}\label{eq:cnK4},
  \end{align}
\item the intrinsic connection 
  \begin{multline}
    0 = 2 \del_{(A}{}^{E} \Gamma_{B)ECD} + \2 K K_{ABCD}
      - \2 K_{AB}{}^{EF} K_{CDEF} \\
      - \frac14 \eps_{C(A}\eps_{B)D}
      \left( K_{EFGH} K^{EFGH} - K^2 \right)
      + 2 \Gamma^E{}_{(A}{}^F{}_{|C|} \Gamma_{B)EDF} \\
      - \frac13 \eps_{C(A}\eps_{B)D} \left( \phi + 6 \Lambda\right) 
      + \Omega E_{ABCD} + \phi_{ABCD},\label{eq:cnGamma}
  \end{multline}
\item the Ricci curvature components
  \begin{align}
    0 &= \del_{(A}{}^{E} \phi_{BCD)E} - \2 K_{(ABC}{}^E
          \phi_{D)E} - \frac{i}{2} B_{ABCD} - \Sigma_{(A}{}^E
          E_{BCD)E},\label{eq:cnphi4}\\
    0 &= \del^{CD} \phi_{ABCD} + \frac13 \del_{AB} \phi + \2 K_{ABCD}
          \phi^{CD} - \2 K \phi_{AB} \nonumber \\
          &+ \Sigma^{CD}E_{ABCD} + 2 \del_{AB} \Lambda, \label{eq:cnphi2}\\
    0 &= \del_{(A}{}^{E} \phi_{B)E} + K_{(A}{}^{CDE}
          \phi_{B)CDE} - i \Sigma^{CD}B_{ABCD}.\label{eq:cnphi}
  \end{align}
\item and the constraints for the Weyl spinor
  \begin{align}
    0 &= \del^{CD} E_{ABCD} - i K_{(A}{}^{CDE} B_{B)CDE},\label{eq:cnE4} \\
    0 &= \del^{CD} B_{ABCD} + i K_{(A}{}^{CDE} E_{B)CDE}.\label{eq:cnB4}
  \end{align}
\item The rest of the constraints is concerned with the conformal
  factor and its derivatives
  \begin{align}
    0 &=\del_{AB} \Omega - \Sigma_{AB},\label{eq:cnOmega} \\
    0 &= \del_{AB} \Sigma + K_{ABCD} \Sigma^{CD} + \Omega
    \phi_{AB}, \label{eq:cnSigma}\\
    0 &= \del_{AB} \Sigma_{CD} - \2 K_{ABCD} \Sigma + \Omega
    \phi_{ABCD} + \frac16 \eps_{C(A}\eps_{B)D} \left( \Omega \phi + 6
      S \right),\label{eq:cnSigma2} \\
    0 &= \del_{AB} S + \phi_{ABCD} \Sigma^{CD} + \2 \phi_{AB} \Sigma 
    - \frac16 \phi \Sigma_{AB} - \Omega \del_{AB} \Lambda + 2 \Lambda
    \Sigma_{AB}.\label{eq:cnS}
  \end{align}
\item The final constraint is the algebraic condition mentioned
  earlier
  \begin{equation}
    \label{eq:algcond}
   0 = 2\Omega S-2\Omega ^2\Lambda -\frac 12\Sigma ^2-\Sigma _{AB}\Sigma ^{AB}.
  \end{equation}
\end{itemize}

Finally, a note on the name conventions of the various spinor
components. In the above equations all the spinor fields are
irreducible except for the extrinsic curvature and the intrinsic three
connection which have the decompositions
\begin{gather}
  K_{ABCD} = K_{4ABCD} - \frac13 \eps_{C(A} \eps_{B)D} K,\\
  \Gamma_{ABCD} = \gamma_{4ABCD} + \2 \eps_{A(C} \gamma_{2D)B} 
  + \2 \eps_{B(C} \gamma_{2D)A} - \frac13 \eps_{C(A} \eps_{B)D} \gamma\\
\end{gather}
with $K_{4ABCD}$ being totally symmetric. The components of the
irreducible parts with respect $(o_A,\iota_A)$ are defined as follows
for a four index spinor $\alpha_{ABCD}$
\begin{multline}
  \alpha_{ABCD} =  \alpha_4\; o_A o_B o_C o_D 
                -4 \alpha_3\; o_{(A} o_B o_C \iota_{D)} \\
                +6 \alpha_2\; o_{(A} o_B \iota_C \iota_{D)}
                -4 \alpha_1\; o_{(A} \iota_B \iota_C \iota_{D)}
                 + \alpha_0\; \iota_A \iota_B \iota_C \iota_D,
\end{multline}
and for a two index spinor
\begin{equation}
  \alpha_{AB} = \alpha_2\; o_A o_B
                -2 \alpha_1\; o_{(A} \iota_{B)}
                + \alpha_0\; \iota_A \iota_B.
\end{equation}
When we have spinors with the same kernel symbol but different numbers
of indices like $\phi_{AB}$ and $\phi_{ABCD}$ we specify the number of
indices in the name of the components, thus obtaining e.g., $\phi_{40}$
as a component of $\phi_{ABCD}$ and $\phi_{22}$ as a component of
$\phi_{AB}$.

\section{The symmetry reduction}
\label{sec:symm}

Finally we discuss a simplification which has been used to reduce the
resource requirements. Since the main interests in this project is the
development of procedures to locate $\scri$, to extract the radiation
information from there and to study several various gauge choices it
is legitimate to assume the existence of a continuous symmetry. This
is because to locate $\scri$ as the zero-set of the conformal factor
is not much more difficult in three dimensions than it is in two. On
the other hand, one needs to have at least two non trivial spatial
dimensions because otherwise the Weyl curvature vanishes identically
so that there will be no radiation present.  Hence, in the sequel we
will assume the existence of a space-like Killing field $\xi^a$ in the
physical space-time, which in addition is required to be hypersurface
orthogonal. In order to exclude numerical problems with coordinate
singularities we follow B.~G.~Schmidt \cite{Schmidt-1996}, who
suggested to look at situations where the Killing vector has no
singular points. This excludes the physically intuitive axisymmetry
which has fixed points on the axis and leaves us with a completely non
physical toy problem. It also has the implication that $\scri$ no
longer has spherical cross sections but toroidal ones. These global
questions are not relevant to local considerations like the influence
of different choices of gauge functions on the solution or even the
question of stability of the numerical method. They do, however,
forcefully come to the fore when it comes to defining and interpreting
global quantities like the Bondi mass or the radiation flux. These are
issues that have not yet been discussed. We will consider them in
somewhat more detail in a later section.

We adapt the gauge to the symmetry. First of all, the Killing vector
in the physical space-time becomes a conformal Killing vector in the
unphysical space-time. Choosing an appropriate conformal gauge we may
achieve that it becomes a Killing vector. Then the conformal factor is
invariant under the symmetry and the conformal gauge freedom is
reduced to multiplication with functions which are also invariant. We
assume that the foliation of $M$ into space-like hypersurfaces is
compatible with the symmetry, i.e., that $\xi^a$ is everywhere tangent
to the hypersurfaces (i.e., orthogonal to $t^a$).  Next we choose the
frame in such a way that one of the space-like legs points along the
Killing vector and that it is invariant under the action of the
symmetry group. Note, that this restricts the available frame rotation
from the orthogonal group to rotations around the Killing
vector. Finally, we take one of the coordinates, say $x^3$ to be the
coordinate along the integral curves of the Killing vector. Then all
components of the geometric quantities with respect to the adapted
frame are independent of that coordinate.

These choices can be made irrespective of whether the Killing vector
is hypersurface orthogonal or not. They do not entail much
simplification in the equations except for the fact that some frame
components and connection coefficients vanish. In particular, the Weyl
curvature still has all of its ten components. However, assuming
hypersurface orthogonality simplifies things considerably. This is
because it is equivalent to the existence of a discrete symmetry
$\xi^a \mapsto -\xi^a$. To explain the simplification it is best to
consider an example. Let $C_{abcd}$ be the Weyl tensor on $M$. The
electric part with respect to the Killing vector $\xi^a$ is
proportional to $\xi^a\xi^cC_{abcd}$ which is symmetric under the
discrete symmetry. However, the magnetic part with respect to $\xi^a$
which is proportional to $\xi^a\xi^cC_{abcd}^\star$ changes sign under
the symmetry hence it has to vanish. This reduces the Weyl curvature
down to five independent components. Similar consideration can be made
for the geometric quantities showing that the number of independent
variables reduces from fifty-three down to thirty-three.

In the space spinor formalism we take the Killing vector to be of the
form $\xi^{AB} = \xi^{BA} \propto o^{(A}\iota^{B)}$. Then the discrete
symmetry implies that the components of almost all of the spinor fields
vanish if they are obtained by contraction with an odd number of $o^A$
and $\iota^A$. Only the fields $F_{AB}$, $\Gamma_{ABCD}$ and
$B_{ABCD}$ have a different behaviour. For them it is the even
components which have to vanish. So, e.g., the Weyl curvature is
described by the five non vanishing components $E_0$, $E_2$, $E_4$, $B_1$
and $B_3$ in agreement with the argument given above.

\section{Conclusion}

The purpose of the present article was the presentation
of the conformal field equations in the space spinor formalism. This
paper is intended to serve as a reference for future articles which
are intended to discuss the numerical implementation of this set of
equations for solving the hyperboloidal initial value problem. We
have discussed the simplification obtained from assuming the existence
of a hypersurface orthogonal Killing vector field.

\section*{Acknowledgments}

Much of this work was done at the Max-Planck-Institut f\"ur
Gravitationsphysik in Potsdam. I wish to thank all the members of the
Mathematical Relativity group there. In particular, discussions with
H. Friedrich, B. Schmidt and P. H\"ubner have been extremely helpful.

\end{document}